\documentclass[aps,prl,reprint,footinbib,superscriptaddress]{revtex4-1}
\usepackage{amsfonts,amssymb,amsmath}
\usepackage{graphicx,epstopdf}
\usepackage{array}
\usepackage{placeins}
\usepackage{slashed}
\usepackage{physics}
\allowdisplaybreaks

\usepackage[color,final]{showkeys} 
\definecolor{refkey}{gray}{0.75}
\definecolor{labelkey}{RGB}{155,48,48}
\renewcommand*\showkeyslabelformat[1]{%
  \fbox{\parbox[t]{0.8\marginparwidth}{\raggedright\normalfont\scriptsize\url{#1}}}}

\usepackage{IEEEtrantools}
\newcommand{\beq}{\begin{equation}}
\newcommand{\eeq}{\end{equation}}
\newcommand{\nn}{\nonumber}

\newcommand{\mD}{\mathcal{D}}

\newcommand{\mN}{\mathcal{N}}

\newcommand{\mS}{\mathcal{S}}

\newcommand{\p}{\partial}

\newcommand{\f}{\frac}

\newcommand{\al}{\alpha}
\newcommand{\be}{\beta}
\newcommand{\ga}{\gamma}         
        
\newcommand{\ep}{\epsilon}

\newcommand{\te}{\theta}

\newcommand{\ka}{\kappa}
\newcommand{\la}{\lambda}       

\newcommand{\rh}{\rho}

\newcommand{\ph}{\phi}          \newcommand{\Ph}{\Phi}
\newcommand{\ps}{\psi}

\newcommand{\bph}{\bar{\ph}}
\newcommand{\bPh}{\bar{\Ph}}

\newcommand{\bps}{\bar{\psi}}

\newcommand{\lan}{\langle}
\newcommand{\ran}{\rangle}

\newcommand{\mO}{\mathcal{O}}

\begin{document}
\title{All tree level scattering amplitudes in Chern-Simons theories with fundamental matter}
\author{Karthik Inbasekar}
\email{karthikin@tauex.tau.ac.il}
\affiliation{Faculty of Exact Sciences, School of Physics and Astronomy, Tel Aviv University, Ramat Aviv 69978, Israel}
\author{Sachin Jain}
\email{sachin.jain@iiserpune.ac.in}
\affiliation{Indian Institute of Science Education and Research, Homi Bhabha Rd, Pashan, Pune 411 008, India}
\author{Pranjal Nayak}
\email{nayak.pranjal@gmail.com}
\affiliation{Department of Theoretical Physics, Tata Institute of Fundamental Research, Navy Nagar, Mumbai 400005, India\\
and\\
Department of Physics \& Astronomy, 265 Chemistry-Physics Building, University of Kentucky, Lexington, 40506, USA}
\author{V Umesh}
\email{vumesh.physics@gmail.com}
\affiliation{National Institute for Theoretical Physics, School of Physics and Mandelstam Institute for Theoretical Physics, University of the Witwatersrand, Johannesburg Wits 2050, South Africa}

\begin{flushleft}
TAUP-3026/17, TIFR/TH/17-30
\end{flushleft}
\begin{abstract}
{We show that Britto-Cachazo-Feng-Witten (BCFW) recursion relations can be used to compute all tree level scattering 
amplitudes in terms of $2\rightarrow2$ scattering amplitude in $U(N)$ ${\mathcal N}=2$ Chern-Simons 
(CS) theory coupled to matter in fundamental representation. As a byproduct, we also obtain a recursion 
relation for the CS theory coupled to regular fermions, even though in this case standard BCFW deformations 
do not have a good asymptotic behaviour.   Moreover at large $N$, $2\rightarrow 2$ scattering can be computed
exactly to all orders in 't Hooft coupling as was done in earlier works by some of the authors. In particular,
for  ${\mathcal N}=2$ theory, it was shown that $2\rightarrow 2$ scattering is tree level exact to all orders 
except in the anyonic channel \cite{Inbasekar:2015tsa}, where it gets renormalized by a simple function of 't Hooft coupling. This suggests that it may be possible to compute the all loop exact result for arbitrary higher point scattering amplitudes at large $N$.}
\end{abstract}
  \maketitle  
\newpage
\vspace*{-5mm}
\section{Introduction}\label{intro}
Chern-Simons (CS) gauge theories coupled to matter fields have a wide variety of applications in areas as diverse as
quantum hall physics,  anyonic physics, topology of three manifolds,  quantum gravity via the AdS/CFT correspondence, etc.
In particular, CS theories coupled to matter in fundamental representation \cite{Aharony:2011jz,Giombi:2011kc} are conjectured to enjoy a strong weak duality, 
which follows from the study of their corresponding bulk duals \cite{Klebanov:2002ja,Sezgin:2002rt,Giombi:2009wh,Giombi:2011kc}.
Moreover, at large N, these theories are exactly solvable \cite{Aharony:2011jz,Giombi:2011kc,Maldacena:2011jn,Maldacena:2012sf}.
This led to impressive large $N,\ka$ (keeping the 't Hooft coupling $\la=\f{N}{\ka}$ fixed) computations to all orders in the 't Hooft coupling in both sides of duality and hence verifying the duality quite convincingly. These computations include exact multipoint current correlators
\cite{Aharony:2012nh,GurAri:2012is,Gurucharan:2014cva,Bedhotiya:2015uga,xyz,Geracie:2015drf,Gur-Ari:2016xff}, exact partition function
\cite{Giombi:2011kc,Jain:2012qi,Yokoyama:2012fa,Aharony:2012ns,Jain:2013py,Takimi:2013zca,Jain:2013gza,
Yokoyama:2013pxa,Minwalla:2015sca,Nosaka:2017ohr}, and exact S-matrices
\cite{Jain:2014nza,Dandekar:2014era,Inbasekar:2015tsa,Yokoyama:2016sbx}. See also
\cite{xyz,Giombi:2016zwa,
Wadia:2016zpd,Radicevic:2016wqn,Giombi:2017txg} for further checks of duality. Recently, the duality was made more precise in
\cite{Radicevic:2015yla,Aharony:2015mjs,Seiberg:2016gmd,Karch:2016sxi} and subsequently generalized to finite N in  
\cite{Karch:2016aux,Hsin:2016blu,Aharony:2016jvv,Benini:2017dus,Gaiotto:2017tne,Jensen:2017dso,Jensen:2017xbs}. 
An example of the strong-weak duality is the duality between CS gauge theory coupled to fundamental 
fermions and CS gauge theory coupled to fundamental critical bosons. Other examples include self dual 
theories, such as ${\mathcal N}=1$, ${\mathcal N}=2$ supersymmetric CS matter theories. At large $N$, 
it was demonstrated that the S matrix for the $2\to2$ scattering computed exactly to all orders in the 't Hooft 
coupling displays an unusual modified crossing relation \cite{Jain:2014nza,Inbasekar:2015tsa,Yokoyama:2016sbx}. 
Moreover, for ${\mathcal N}=2$ theory, the result is tree level exact \cite{Inbasekar:2015tsa} except in the 
anyonic channel, where it gets renormalized by a simple function of the 't Hooft coupling. 

A natural question to ask would be, is it possible to compute  arbitrary $m\rightarrow n$ scattering amplitudes at
all values of the 't Hooft coupling at large $N,\ka$? Given the simplicity of the results at least in the 
supersymmetric case, it is also interesting to ask if the computability of scattering amplitudes extends to finite $N,\ka$.
As a first step towards these questions, we compute all tree level amplitudes for  the ${\mathcal N}=2$ theory and 
the regular fermionic theory. We show that a $m\rightarrow n$ scattering amplitude can be computed recursively in terms of the $2\to2$ scattering amplitudes in these theories. Similar recursion relations in three dimensions
were first developed in \cite{Gang:2010gy}, in the context of the Aharony-Bergman-Jafferis-Maldacena (ABJM) theory and
subsequently applied to other theories like 3d super-Yang-Mills in \cite{Lipstein:2012kd}
 and massive 3d $\mN=2$ gauge theories in \cite{Agarwal:2013tpa}. Note that, the self-dual $\mN=2$ supersymmetric theory is particularly interesting and important since via RG flow, we can obtain non supersymmetric dual pairs such as critical bosons coupled to CS and regular fermions coupled to Chern-Simons \cite{Jain:2013gza,xyz}.

\vspace*{-5mm}
\section{Four point scattering amplitude}\label{4ptsca}
In this Letter, we compute scattering amplitudes in fermion coupled to $SU(N)$ CS theory (FCS) 
\begin{equation}\label{fcs}
 \int d^3x 
\biggl[-\f{\ka}{4\pi}\ep^{\mu\nu\rh}\text{Tr}\left( A_\mu\p_\mu A_\rh-\f{2i}{3}A_\mu A_\nu
A_\rh  \right)+\bps i \slashed{\mD} \ps\biggr]\ ,
\end{equation}
and in ${\mN=2}$ CS matter theory coupled to a Chiral multiplet given by
\begin{align}\label{susycs}
& \mS_{\mN=2}^L = \int d^3x 
\biggl[-\f{\ka}{4\pi}\ep^{\mu\nu\rh}\text{Tr}\left( A_\mu\p_\mu A_\rh-\f{2i}{3}A_\mu A_\nu
A_\rh  \right)\nonumber \\
&\quad +\bps i \slashed{\mD} \ps-\mD^\mu\bph\mD_\mu\ph+\f{4\pi^2}{\ka^2}(\bph\ph)^3
+\f{4\pi}{\ka}(\bph\ph)(\bps\ps)\nn\\
&+\f{2\pi}{\ka}(\bps\ph)(\bph\ps)\biggr]\ .
\end{align}
For our purposes, it is convenient to introduce the spinor helicity basis \cite{Elvang:2015rqa} defined by
\begin{equation}
p_i^{\alpha\beta}=p_i^{\mu}\sigma_{\mu}^{\alpha\beta}=\lambda_i^{\alpha}\lambda_i^{\beta},~~ (p_i+p_j)^2= 2 p_i.p_j=\langle \lambda_i^{\alpha} \lambda_{i,\alpha}\rangle^2\ .
\end{equation}
Below we use the notation $ \langle\lambda_i^{\alpha} \lambda_{j,\alpha}\rangle= \langle i j\rangle.$
For a supersymmetric amplitude, the standard procedure involves introduction of on-shell grassman variables $\theta$ such that the super-creation and super-annihilation operators are given by
\begin{equation}\label{eq:super-cr-an}
A_i= a_i+ \theta_i \alpha_i ,~~A_i^{\dagger}=\theta_i a^{\dagger}_i+  \alpha^{\dagger}_i ,
\end{equation}
where $\left(a^{\dagger}_i,a_i\right)$/$\left(\alpha^{\dagger}_i,\alpha_i\right)$  create and annihilate a boson/fermion with momenta $p_i$ respectively.
The two on-shell supercharges for $n$ point scattering amplitudes  are given by
\begin{equation}\label{SChrg}
Q=\sum_{i=1}^{n}q_i =\sum_{i=1}^{n}\lambda_i \theta_i,~~ {\bar Q}=\sum_{i=1}^{n}{\bar q}_i =\sum_{i=1}^{n}\lambda_i \partial_{\theta_i}\ .
\end{equation}

For FCS theory in \eqref{fcs}, the tree level $2\to 2$ scattering amplitude is given by \cite{Jain:2014nza}
\begin{equation}\label{fer4pt}
A_4^{F}=\langle\bps(p_1) \ps(p_2) \bps(p_3) \ps(p_4) \rangle= \frac{\langle 1 2\rangle \langle 2 4\rangle}{\langle 2 3\rangle}\delta(\sum_{i=1}^{4}p_i)\ .
\end{equation}
For ${\mN=2}$ theory in \eqref{susycs}, the tree level $2\to 2$ super amplitude is given by 
\begin{equation}\label{eq:super4pt}
A_4^{S}=\frac{\langle 1 2\rangle}{\langle 2 3\rangle}\delta(\sum_{i=1}^{4}p_i)Q^2=\frac{\langle 1 2\rangle}{\langle 23\rangle}\delta(\sum_{i=1}^{4}p_i)\sum_{1=i<j}^{4}\langle i j\rangle \theta_i \theta_j\ .
\end{equation}
Here $A_4^{S}$ is the super-amplitude computed using the super-creation/annihilation operators defined in \eqref{eq:super-cr-an}. Any component amplitude can be obtained from \eqref{eq:super4pt} by picking up the coefficient of products of two $\theta$'s. 

\section{Higher point scattering amplitude}\label{higherpt}
BCFW recursion relations are an efficient method to compute and express arbitrary higher point scattering amplitudes in terms of product of lower point amplitudes. Standard procedure for BCFW involves the deformation of two external momenta of the particles by a complex parameter $z$ such that the particles continue to remain ``on shell'' and the total momentum conservation of the process continues to hold. In $3D$, BCFW deformations are a little different than in $4D$ and were first discussed in \cite{Gang:2010gy} (We follow their notations closely).  BCFW recursion relations are applicable in $3D$ provided that the higher point amplitudes are regular functions at both $z\to\infty$ and $z\to0$. In the following section we study the $z\to\infty$ (and $z\rightarrow 0$) behavior of the amplitudes in the theories described earlier. We find it convenient to deform color contracted (we label them as \lq$1$\rq and \lq$2$\rq) external legs. In 3-dimensions, momentum deformation of particles 1 and 2 can be written in terms of the spinor-helicity variables as
\begin{equation}\label{BCFWd}
\left(\begin{array}{c} {\hat \lambda}_1 \\ {\hat \lambda}_2 \end{array}\right) = R \left(\begin{array}{c} \lambda_1 \\  \lambda_2 \end{array}\right),~~\rm{where}~~~
R=\left(\begin{array}{cc} \frac{z+z^{-1}}{2}& -\frac{z-z^{-1}}{2~i}\\ \frac{z-z^{-1}}{2~i} & \frac{z+z^{-1}}{2} \end{array}\right)\ .
\end{equation}

In the theories \eqref{fcs},\eqref{susycs}, all 3-point vertices involve gauge fields and since the CS gauge field does not have an on shell propagating degree of freedom, it follows that only even-point functions are non-zero. This also implies that the 4-point functions are fundamental building blocks for higher point functions.

Under the deformation \eqref{BCFWd}, any tree-level scattering amplitude for FCS in \eqref{fcs} is not well behaved at large $z$ and hence doesn't obey the requirements of BCFW. However this situation is cured for the ${\mN=2}$ theory defined in  \eqref{susycs}. Additionally, conservation of the super-charges in \eqref{SChrg} require that the on-shell spinor variables $\theta$ be deformed as
\begin{equation}\label{BCFWd-th}
\left(\begin{array}{c} {\hat \theta}_1 \\ {\hat \theta}_2 \end{array}\right) = R \left(\begin{array}{c} \theta_1 \\  \theta_2 \end{array}\right)\ ,
\end{equation} where the $R$ matrix is defined by \eqref{BCFWd}.

Let us denote the $2n-$point super-amplitude as $A_{2n}(\lambda_1,\lambda_2,\cdots\lambda_{2n},\theta_1,\theta_2,\cdots\theta_{2n})$ and the deformed amplitude by 
$A_{2n}({\hat \lambda}_1,{\hat \lambda}_2,\cdots\lambda_{2n},{\hat \theta}_1,{\hat \theta}_2,\cdots\theta_{2n},z)$. The deformed super-amplitude can be explicitly written as an expansion in the $\te$ variables as follows
\begin{align}\label{BCFWsuper}
A_{2n}(z)&= A^{0}(z) + A^{1}(z) {\hat \theta}_1(z) +  A^{2}(z) {\hat \theta}_2(z) \nonumber \\
& \hspace{3cm}+ A^{12}(z) {\hat \theta}_1(z) {\hat \theta}_2(z) \nn\\
&=A^{0}(z) + {\tilde A}^{1}(z)  \theta_1 +  {\tilde A}^{2}(z)  \theta_2 + A^{12}(z)  \theta_1  \theta_2 ,
\end{align}
where in the last line of \eqref{BCFWsuper} we have used \eqref{BCFWd} and the  fact that $ {\hat \theta}_1(z) {\hat \theta}_2(z) = \theta_1 \theta_2$. We have also defined 
\begin{align}
\begin{pmatrix}
\tilde{A}^1(z)\\ 
\tilde{A}^2(z)
\end{pmatrix}
= 
R^T\begin{pmatrix}
A^1(z)\\ 
A^2(z)\end{pmatrix} \ ,
\end{align}
where $R^T$ is the transpose of the R matrix defined in \eqref{BCFWd}, with $RR^T=1$.
The super-momentum conservation implies  that the large $z$ behavior of the super-amplitude $A_{2n}(z)$ is identical to that of the components $A^{0}, A^{12}$. Hence it is sufficient to show that either of $A^{0}$ or $A^{12}$ are well behaved since supersymmetric ward identity guarantees the required behavior for the rest of the amplitudes. It is convenient to write the fields in pair wise contractions since they transform in the fundamental representation of the gauge group. For instance we are interested in the large z behavior of amplitudes such as  $(\bps_1^i\ph_{2i})(\bph_3^j\ps_{4j})\ldots$  and  $(\bph_1^i\ps_{2i})(\bps_3^j\ph_{4j})\ldots$, where $\ldots$ represent color contracted bosonic or fermionic particles allowed by interactions in \eqref{susycs}. These amplitudes appear in $A^0, A^{12}$ in \eqref{BCFWsuper} respectively.

 We have checked explicitly by Feynman diagrams that the amplitude $A^0=A_6(\bps_1\ph_2\bph_3\ps_4\bph_5\ph_6)$ is well behaved. We discuss the large $z$ behavior of the general $2n$ point amplitude using the background field method \cite{ArkaniHamed:2008yf} in the next section.

\section{Asymptotic behavior of amplitudes}\label{bacg}

\begin{figure}[t!]
\includegraphics[scale=0.45]{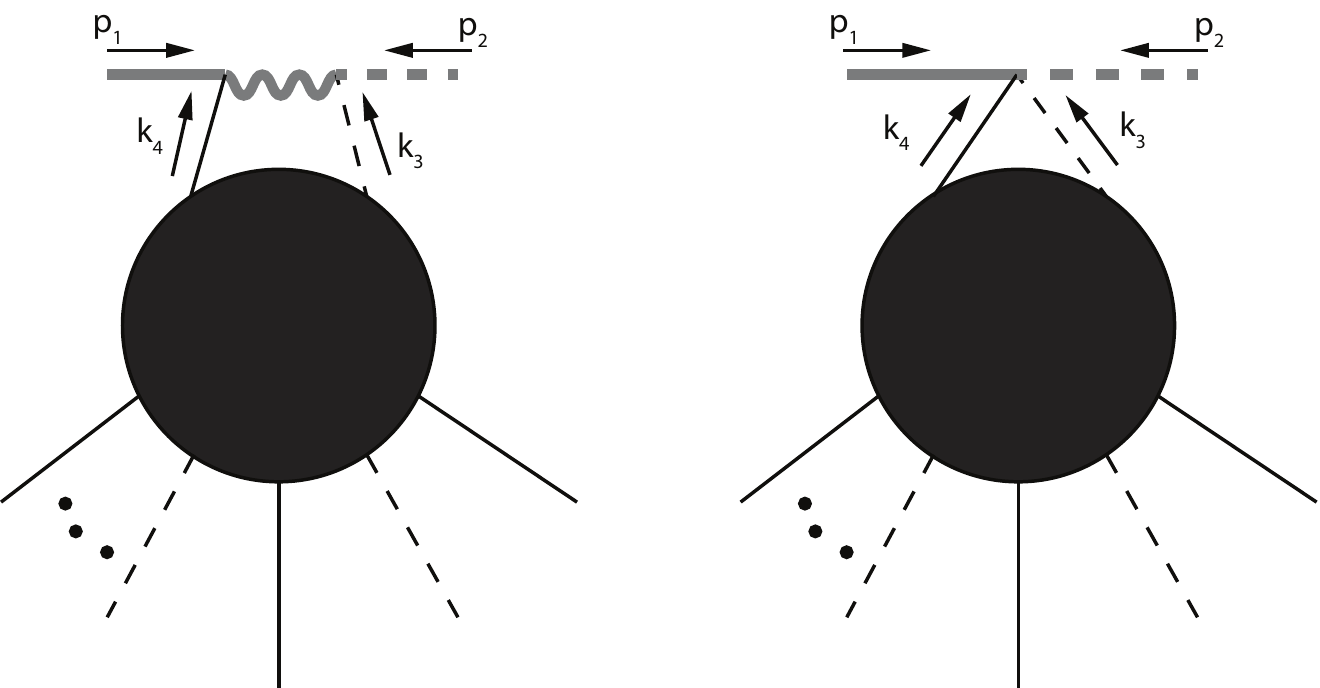}
\caption{The diagrams that have a non-regular $z\to\infty$ behavior. $\mathcal O(z)$ part of these two diagrams cancel against each other to give a regular $z\to\infty$ behavior of the total amplitude. In the above diagram, the solid lines correspond to fermions and the dashed lines correspond to bosons. This amplitude appears in $A^{0}$ in \eqref{BCFWsuper}. The gray coloured lines corresponds to deformed hard particle.}
\label{fig:diagcancel}
\end{figure}

To understand the  large $z$ behavior of various scattering amplitudes,  it is extremely useful to think from the background field method point of view introduced in \cite{ArkaniHamed:2008yf}. Here $z$-deformed particles are considered as hard particles propagating in a background of soft particles. The amplitude is modified due to (a) modified propagator of intermediate hard particle; (b) the modified contribution of various vertices; and, (c) modified fermion wave function, in case an external deformed particle is a fermion.  Detailed analysis shows (we follow closely \cite{ArkaniHamed:2008yf,Gang:2010gy}) that the non-trivial $z\to\infty$ behavior of the amplitude is due to diagrams of the kind depicted in fig. \ref{fig:diagcancel}. 
The values of these diagrams are:
\begin{align}
	&\!\!\!\text{Gauge-field exchange: } \frac { 4\pi i}{ \ka } \lan k_4|\ga^\mu|1 \ran \frac {k_3^\nu p_2^\rho }{ (k_3+p_2)^2 } \ep_{\mu\nu\rho}\label{gvtx}\\
	&\text{Contact vertex: } -\frac{ 2\pi }{ \ka } \lan k_4|1 \ran \label{eq:cont-vert}
\end{align}
Under the 1-2 $z$-deformations, \eqref{BCFWd}, in the $z\to\infty$ limit the $\mO(z)$ part of the amplitude cancels and the amplitude behaves as $\mO(1/z)$. Hence this amplitude has a regular $z\to\infty$ behavior for ${\mathcal N}=2$ theory. This cancellation works even for the 4-point function $\lan \bps_1 \ph_2 \bph_3 \ps_4 \ran$, which receives contributions from the diagrams in fig. \ref{fig:diagcancel} with the blob removed and $k_3\to p_3, k_4\to p_4$ are taken to be on-shell momenta. It is important to emphasize that we need minimum ${\mathcal N}=2$ amount of supersymmetry for this to work \footnote{For instance, for $\mN=1$ theory, the Lagrangian for which can be found in \cite{Inbasekar:2015tsa} (Equation (2.11)), \eqref{eq:cont-vert} is modified to $(-2\pi w/\kappa)\,{\langle41\rangle}$ whereas \eqref{gvtx} remains the same. Here $w$ is a free parameter in $\mN=1$ theory. This implies that only at $w = 1$, the $N = 1$ theory has a good large z behavior. This is exactly the point in the $w$ line where the supersymmetry of the theory gets enhanced to $\mN = 2$.}.
\section{Recursion relations in $\mN=2$ theory}\label{recrel}

\begin{figure}[ht!]
\begin{center}
\includegraphics[scale=0.25, trim=0cm 0cm 0cm -2cm, clip]{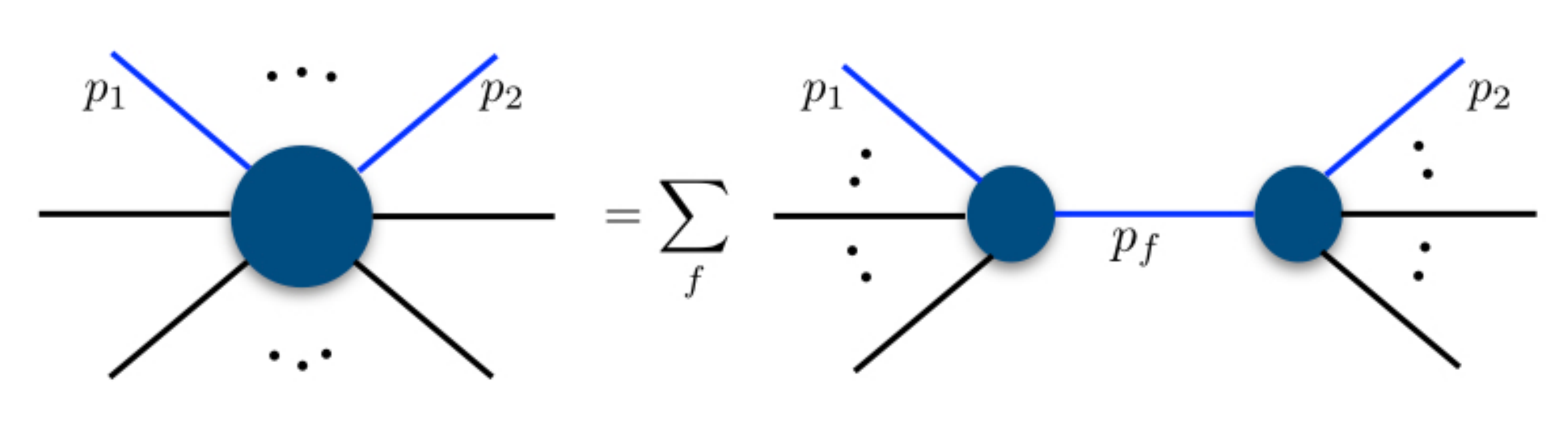}
  \caption{Recursion formula for a 2n point amplitude: The black lines denote the undeformed legs, the external gray lines represent the deformed legs and $p_f$ represents the momentum in the factorization channel.}\label{figx}
 \end{center}
\end{figure}
\vspace{0.2cm}
In the last section, we have demonstrated that $A^{0}$ is well behaved in large $z$. Hence we can apply the BCFW recursion relation directly to the super amplitude in the left hand side of \eqref{BCFWsuper}. The recursion formula for a $2n$ point superamplitude can be expressed in terms of lower point superamplitudes as follows (see fig \ref{figx})
\begin{widetext}
\begin{align}\label{bcfw}
A_{2n}(z=1)=\sum_f \int  \f{d \te}{p_f^2}\biggl(z_{a;f} \f{z_{b;f}^2-1}{z_{a;f}^2-z_{b;f}^2} A_L(z_{a;f},\te)A_R(z_{a;f},i\te)  +(z_{a;f}\leftrightarrow z_{b;f}) \biggr),\\
\label{residue} \left(z_{a;f}^2~,~z_{b;f}^2\right) = \f{-(p_f-p_2).(p_{f}+p_1)\pm\sqrt{(p_f-p_2)^2 (p_{f}+p_1)^2}}{4 q.(p_f-p_2)}\ ,
\end{align}
\end{widetext}
where the integration is over the intermediate Grassmann variable $\te$ and $A_{2n}(z=1)$ is the undeformed $2n$-point amplitude.
In the above, $p_f$ is the undeformed momentum that runs in the factorization channel $f$ and the summation in \eqref{bcfw} runs over all the factorization channels corresponding to different intermediate particles going on-shell. Here, $z_{a;f}$ and $z_{b;f}$ are given by \eqref{residue}, where the null momenta $q$ are defined in terms of the spinor helicity variables as 
\begin{align}\label{nmom}
q^{\al\be} =\f{1}{4} (\la_2+i \la_1)^\al (\la_2+i \la_1)^\be\ .
\end{align}
Note that \eqref{bcfw} has a very similar form (but not quite the same as discussed below) to the one obtained in \cite{Gang:2010gy} for the ABJM theory\footnote{ Although, formula \eqref{bcfw} looks very similar to ABJM case, the details are different since the external matter particles are in fundamental representation. For example, in general there will be more factorization channels here as compared to the ABJM case. For example, in the six point function, as will be clear below, there are two factorized channels, where as for the corresponding deformation in ABJM, there is only one factorized channel.} that enjoys $\mN=6$ supersymmetry. It is remarkable that such recursion formulae exist in a theory with much lesser supersymmetry such as the one in discussion.

Appearance of square roots in the expression \eqref{residue} could be seen as a concern for giving rise to branch-cuts 
in the amplitudes. However, note that $A_L(-z) A_R(-z) = -A_L(z)A_R(z)$ in the integrand of \eqref{bcfw}\footnote{Under $z\to-z$, both $\hat\lambda\to-\hat\lambda$ and $\hat\eta\to-\hat\eta$ (see \eqref{BCFWd} and \eqref{BCFWd-th}). From \eqref{eq:super-cr-an} one can deduce that the scaling of the onshell fields $\Phi,\bar\Phi$ to be $+1/-1$ under $z\to-z$. In our BCFW deformations, we deform one $\Phi$ and one $\bar\Phi$, and therefore, $A_L A_R \to -A_L A_R$. See \cite{Gang:2010gy} for a related discussion in the context of the ABJM theory.}. Also the prefactor is an odd function of $z$. Consequently, the total integrand is an even function of $z_{a;f}$ and $z_{b;f}$ and 
hence only depends on $z^2_{a;f}$ and $z^2_{b;f}$. Moreover, the integrand is also symmetric under
$z_{a;f}\leftrightarrow z_{b;f}$. This implies that the amplitude is only a function of $z_{a;f}^2+z_{b;f}^2$ and 
$z_{a;f}^2 z_{b;f}^2$, and hence there are no square roots in the final expression.

As an explicit demonstration of recursion relations in \eqref{bcfw}, consider the six point\footnote{A general six point super amplitude in $\mN=2$ theory can be written in terms of two independent functions as 
\begin{equation*}\begin{aligned}
A_6&= Q^2 \Bigg(f_1(p) \sum_{i=1}^{3}\epsilon^{ijk}\lambda(p_j)\lambda(p_k)\theta_i \\
&\hspace{2cm}+ f_2(p) \sum_{i=4}^{6}\epsilon^{ilm}\lambda(p_l)\lambda(p_m)\theta_i \Bigg) 
\end{aligned}\end{equation*} where $Q=\sum_{i=1}^{6}\lambda_i\theta_i$ as defined in \eqref{SChrg}.} amplitude $A_6(\la_1\ldots\la_6)\equiv (\bph\ps)(\bps\ph)(\bph\ph)$ in the $\mN=2$ SCS theory. This  amplitude  factorizes into two channels as shown in fig \ref{figz}:

\begin{widetext}

\begin{figure}[h]
\begin{center}
\includegraphics[scale=0.5]{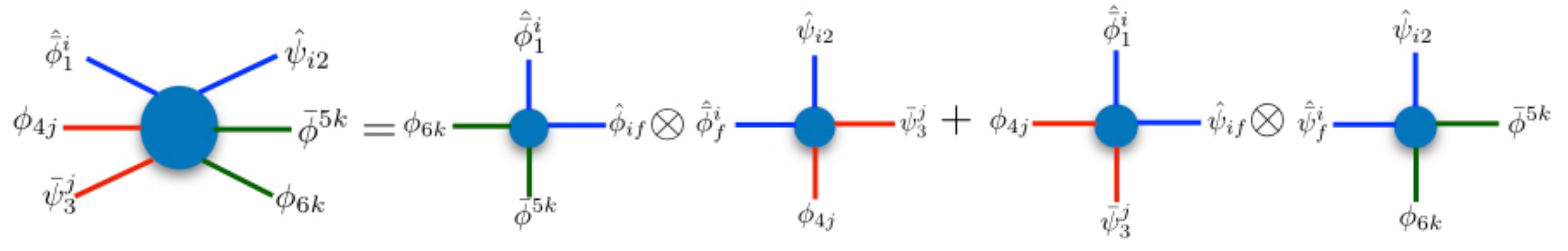}
  \caption{BCFW recursion for the six point amplitude. Factorization into two channels. Each four point amplitude on the RHS is on shell. Two adjacent lines with the same color are color contracted. Note that the gray lines in particular represent the BCFW deformed legs. \vspace{-1cm}}\label{figz}
 \end{center}
\end{figure}
\end{widetext}
The recursion formula can be explicitly written as
\vspace{-0.2cm}
\begin{align}\label{sixpointbcfw}
&\lan\bph_1\ps_2 \bps_3\ph_4 \bph_5\ph_6 \ran=\nn\\
&\biggl(z_{a;f} \f{z_{b;f}^2-1}{z_{a;f}^2-z_{b;f}^2}\lan\hat{\bph}_1\hat{\ph}_f\bph_5\ph_6\ran_{z_{a;f}} \lan\hat{\bph}_{(-f)}\hat{\ps}_2\bps_3\ph_4\ran_{z_{a;f}}  \nn\\
&+(z_{a;f}\leftrightarrow z_{b;f}) \biggr)\f{i}{p^2_{f}}\bigg|_{p_f=p_{234}}\nn\\
&+\biggl(z_{a;f} \f{z_{b;f}^2-1}{z_{a;f}^2-z_{b;f}^2}\lan \hat{\bph}_1\hat{\ps}_f\bps_3\ph_4\ran_{z_{a;f}} \lan \hat{\bps}_{(-f)}\hat{\ps}_2\bph_5\ph_6 \ran_{z_{a;f}} \nn\\
& +(z_{a;f}\leftrightarrow z_{b;f}) \biggr)\f{i}{p^2_{f}}\bigg|_{p_f=p_{256}} ,\\[15pt]
=&\pqty{\frac{ 32\pi^2 i}{\ka^2}}\Bigg[\frac{\lan2|p_4|3\ran \, p_{12}\cdot p_{56} - \lan2|p_1|3\ran \, p_{34}\cdot p_{56}} {p^2_{256}\;p^2_{124}}  \nn\\
&+\Bigg( \lan3|p_{12}|5\ran \, \Big( \lan2|p_1|5\ran \, p_{34}\cdot p_{56} - \lan2|p_6|5\ran \, p_{34}\cdot p_{12} \Big) \nn\\
&\quad-  \lan34\ran \lan12\ran \, \Big( \lan1|p_6|4\ran \, p_{12}\cdot p_{56} - \lan1|p_2|5\ran \, \lan4|p_6|5\ran \Big) \Bigg)\nn\\
& \qquad \qquad \qquad \times \frac1 {p^2_{234}\;p^2_{123}\;p^2_{126}} \Bigg].
\end{align}
Fields with hats corresponds to deformed momenta. We have checked \eqref{sixpointbcfw} explicitly by computing the relevant Feynman diagrams. It is a curious fact that, the total number of Feynman graphs that contribute to $A_6$ is 15. Of these, eleven  are reproduced by the channel $p_f=p_{234}$ and the remaining four in the channel $p_f=p_{256}$. Moreover, we have also reproduced the correct additional poles in the respective channels. The final answer is manifestly free of any spurious poles and square-roots as we argued above.

\section{\label{fermionrec}Recursion Relations in the fermionic Theory}
In this section, we show that the BCFW recursion relations can be used to compute $2n-$point amplitude $A_{2n}=({\bar\psi_1}\psi_2)\cdots ({\bar\psi_{2n-1}}\psi_{2n})$ for the regular fermionic theory coupled to CS gauge field \eqref{fcs}. If we apply \eqref{BCFWd} to this amplitude, it is easy to show that, it does not have a good large $z$ (as well as $z\rightarrow0$) behavior, hence we cannot readily apply the BCFW recursion relation\footnote{There will be some non-trivial boundary terms that do not vanish and in general there are no good prescriptions to compute them systematically.} to determine all higher point fermionic amplitudes. However, we show below that we can use the recursion relation of the ${\mathcal N}=2$ to write a recursion relation for the fermionic theory. 

As a first step towards this, let us note that the Feynman diagrams for any tree-level all-fermion scattering amplitude in the $\mN=2$ theory \eqref{susycs} is identical to that of the tree-level scattering amplitude in the fermionic theory \eqref{fcs}. In the previous section we proved for the $\mN=2$ theory that an arbitrary higher-point super-amplitude can be written only in terms of the 4-point super-amplitude. Same can be said for the component amplitudes including the purely fermionic component amplitude \footnote{Note that the recursion relation in the $\mN=2$ theory \eqref{bcfw} does not directly give $({\bar\psi_1}\psi_2)\cdots ({\bar\psi_{2n-1}}\psi_{2n})$ in terms of the lower point fermionic amplitude. However, we can use BCFW relations recursively to write down any higher point amplitude in terms of four point amplitudes such as $({\bar\psi}\psi)({\bar\psi}\psi)$,  $({\bar\phi}\phi)({\bar\phi}\phi)$, $({\bar\phi}\phi)({\bar\psi}\psi)$, $({\bar\phi}\psi)({\bar\psi}\phi)$ etc. Moreover, at the level of the four point amplitude, one can rewrite this in terms of $({\bar\psi}\psi)({\bar\psi}\psi)$. For example, $({\bar\psi}\psi)({\bar\phi}\phi) =\frac{\langle 23\rangle}{\langle 24\rangle} ({\bar\psi}\psi)({\bar\psi}\psi)$. This implies that we get a recursion relation for $({\bar\psi_1}\psi_2)\cdots ({\bar\psi_{2n-1}}\psi_{2n})$ in terms of lower point fermionic amplitudes only. Hence this can be interpreted as a BCFW recursion relation in the regular fermionic theory coupled to CS gauge field \eqref{fcs}.}. Let us note that for the four point super-amplitude, supersymmetry relates all the component 4-point amplitudes to one component amplitude, which can be taken to be 4-fermion scattering amplitude (see \eqref{eq:super4pt}). Thus an arbitrary higher-point component amplitude can be written only in terms of 4-fermion amplitude. This can be recursively done for an arbitrary $2n$ point amplitude, however for simplicity we write the recursion relation for the six point amplitude below
\begin{align}\label{eq:factor-all-fermi-2}
	& \lan \bps_1 \ps_2 \bps_3 \ps_4 \bps_5 \ps_6 \ran = \nn\\
	&\Biggr( z_{a;f} \f{z_{b;f}^2-1}{z_{a;f}^2-z_{b;f}^2} \qty[ \frac { z_{a;f}^2 + 1 }{ 2 z_{a;f} } + i \; \frac { z_{a;f}^2 - 1 }{ 2 z_{a;f} } \; \frac { \lan \hat14 \ran }{ \lan \hat f 4 \ran } \frac { \lan \hat f 6 \ran }{ \lan \hat26 \ran } ] \nn\\
	& \hspace{3cm} \times \lan {\hat \bps}_{1} {\hat \ps}_{ f} \bps_3 \ps_4 \ran  \lan {\hat\bps}_{(- f)} {\hat\ps}_{2} \bps_5 \ps_6 \ran_{z_{a;f}}\nn \\ 
	&+ (z_{a;f}\leftrightarrow z_{b;f}) \Biggr) \frac 1 {p_f^2}\bigg|_{p_f=p_{256}}\nn\\[5pt]
	+&\Biggr( z_{a;f} \f{z_{b;f}^2-1}{z_{a;f}^2-z_{b;f}^2} \qty[ \frac { z_{a;f}^2 + 1 }{ 2z_{a;f} } + i \; \frac { z_{a;f}^2 - 1 }{ 2z_{a;f} } \; \frac { \lan \hat16 \ran }{ \lan \hat f 6 \ran } \frac { \lan \hat f 4 \ran }{ \lan \hat24 \ran } ] \nn\\
	& \hspace{3cm} \times  \lan  {\hat \bps}_{1} {\hat \ps}_{ f} \bps_5 \ps_6 \ran  \lan {\hat\bps}_{(- f)}{\hat \ps}_{ 2} \bps_3 \ps_4 \ran_{z_{a;f}} \nn\\
	&+ (z_{a;f}\leftrightarrow z_{b;f}) \Biggr) \frac 1 {p_f^2}\bigg|_{p_f=p_{234}}, \nn\\[15pt]
=&\pqty{ \frac{ 16\pi^2 i}{\ka^2}} \pqty{\frac{1}{p_{124}^2 \; p_{125}^2 \; p_{256}^2}-\frac{1}{ p_{123}^2 \; p_{126}^2 \; p_{234}^2}} \times \nn\\
&\Bigg[ -\lan1|p_{34}|2\ran \, \lan3|p_{56}|4\ran \,\lan5|p_{12}|6\ran  + \lan56\ran \lan12\ran \lan34\ran \times \nn\\
& \Big( \lan5|p_{12}|6\ran \, \lan56\ran + \lan3|p_{56}|4\ran \, \lan34\ran + \lan1|p_{34}|2\ran \, \lan12\ran \Big) \Bigg].
\end{align}
The above answer is remarkably simple and is manifestly invariant under the permutations of particle pairs, $\{12\},\{34\}$ and $\{56\}$, as expected.

\section{Discussion}
In this letter we presented recursion relations for all tree level amplitudes in $\mN=2$ CS matter theory and CS theory coupled to regular fermions. Below we discuss some interesting open questions for future research.

It was shown in \cite{Inbasekar:2015tsa}, that the $2\rightarrow 2$ scattering amplitude in the ${\mathcal N}=2$ theory does not get renormalized except in the anyonic channel, where it gets renormalized by a simple function of the 't Hooft coupling. A natural question is, why in the $\mN=2$ theory the scattering amplitude has such a simple form, whereas the corresponding amplitudes in the fermionic \cite{Jain:2014nza} and other less susy ${\mathcal N}=1$ \cite{Inbasekar:2015tsa} theories are quite complicated; and, if the simplicity of the amplitudes continues to persist with higher point amplitudes. It is also interesting to explore an analog of the Aharonov-Bohm phase for higher point amplitudes. It may very well turn out that the Aharanov-Bohm phases of higher point amplitudes are products of the Aharonov-Bohm phases of the $2\to2$ amplitude.  BCFW recursion relations provide a strong indication towards this result.

To answer the above questions, we need to compute higher scattering amplitudes to all orders in $\la$. A possible way is to investigate the Schwinger-Dyson equation.  However, the Schwinger-Dyson equation approach is quite complicated even at the $6$-point level. A refined approach might be to look for a larger class of symmetries such as dual superconformal symmetry \cite{Inbasekar:2017sqp}, Yangian symmetry and use the powerful formulation of \cite{ArkaniHamed:2012nw} to obtain results. Given the fact that, these theories are exactly solvable  at large-$N$ as well as the fact that ${\mathcal N}=2$ theory is self-dual, it could turn out that the ${\mathcal N}=2$ theory may be one of the simplest playing grounds to develop new techniques in computing S-matrices to all orders \cite{ArkaniHamed:2012nw}. 
Furthermore exact solvability at large $N$ indicates that these models might even be integrable. One possible way to investigate integrability is to show the existence of an infinite dimensional Yangian symmetry.  
Since these theories relate to various physical situations, any of the above exercises may provide insight into finite $N,\ka$ computations.

\textbf{Acknowledgements:} We thank O. Aharony, S. Ananth, S.Banerjee,  R. Gopakumar, Nima Arkani Hamed, T. Hartman, Y-t. Huang, S. Kundu, R. Loganayagam, J. Maldacena, G. Mandal, S. Minwalla, S. Mukhi, S. Raju,  A. Sever, T. Sharma, R. Soni, J. Sonnenschein, S. Trivedi, and S. Wadia for helpful discussions. We would like to thank Nima Afkhami-Jeddi and Amirhossein Tajdini for collaboration during the initial stages of the project. We would like to thank Y-t. Huang for sharing a useful mathematica code with us. Special  thanks S. Minwalla for very useful and critical discussions.
The work of KI was supported in part by a center of excellence supported by the Israel Science Foundation (grant number 1989/14), the US-Israel bi-national fund (BSF) grant number 2012383 and the Germany Israel bi-national fund GIF grant number I-244-303.7-2013. S.J. would like to thank TIFR for hospitality at various stages of the work. Some part of the work in this paper was completed while SJ was a postdoc at Cornell and his research  was supported  by grant No:488643 from the Simons Foundation. The work of PN is supported partly by Infosys Endowment for the study of the Quantum Structure of Space Time and Indo-Israel grant of S. Minwalla; and partly by the College of Arts and Sciences of the University of Kentucky. We would also like to thank people of India for their steady support in basic research. 

\bibliographystyle{apsrev4-1}
\bibliography{BCFWPRL.bib,BCFWnotesPRL}

\onecolumngrid
\newpage
\vskip 2cm
\section{Appendix}
\subsection{Large $z$ behavior of the six point scattering amplitude $(\bph(p_1)\ps(p_2))(\bps(p_3)\ph(p_4))(\bph(p_5)\ph(p_6))$}
\begin{figure}
\begin{minipage}{\textwidth}
\includegraphics[width=12cm,height=4cm]{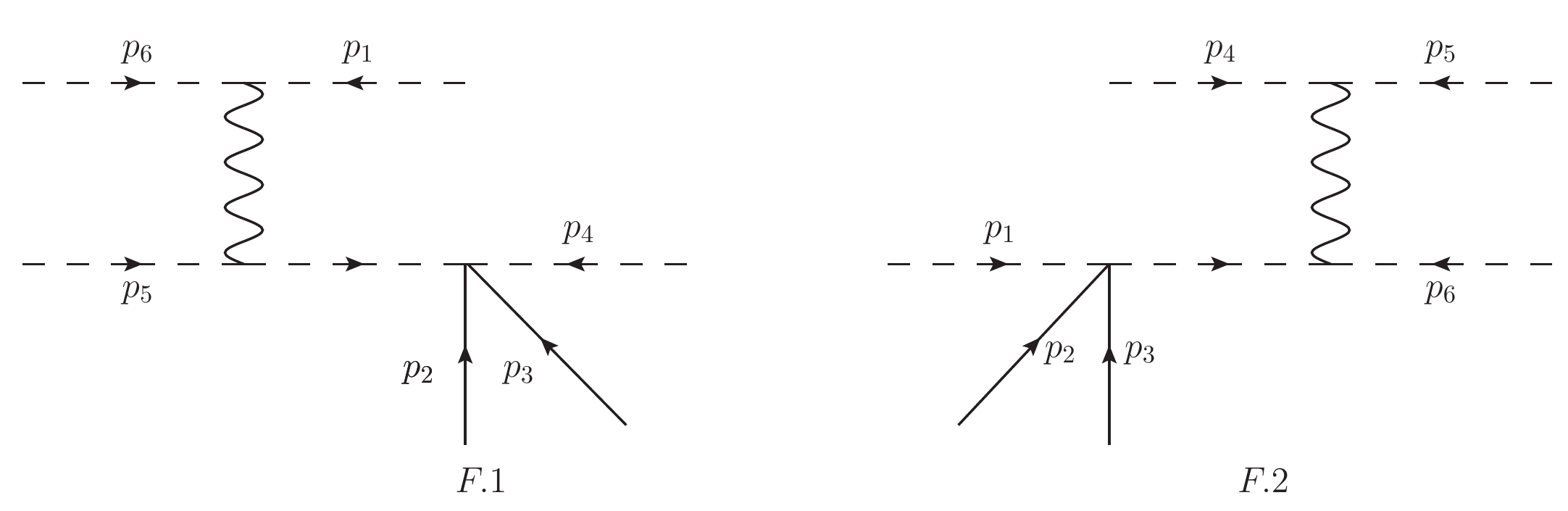}
\end{minipage}
\begin{minipage}{\textwidth}
\includegraphics[width=6cm,height=3.5cm]{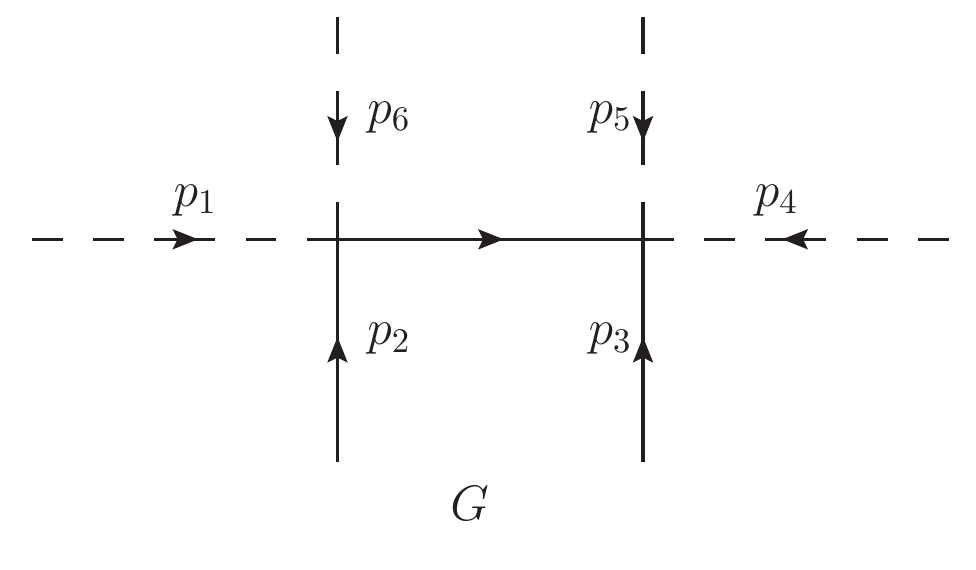}
\end{minipage}
\begin{minipage}{\textwidth}
\includegraphics[width=12cm,height=4cm]{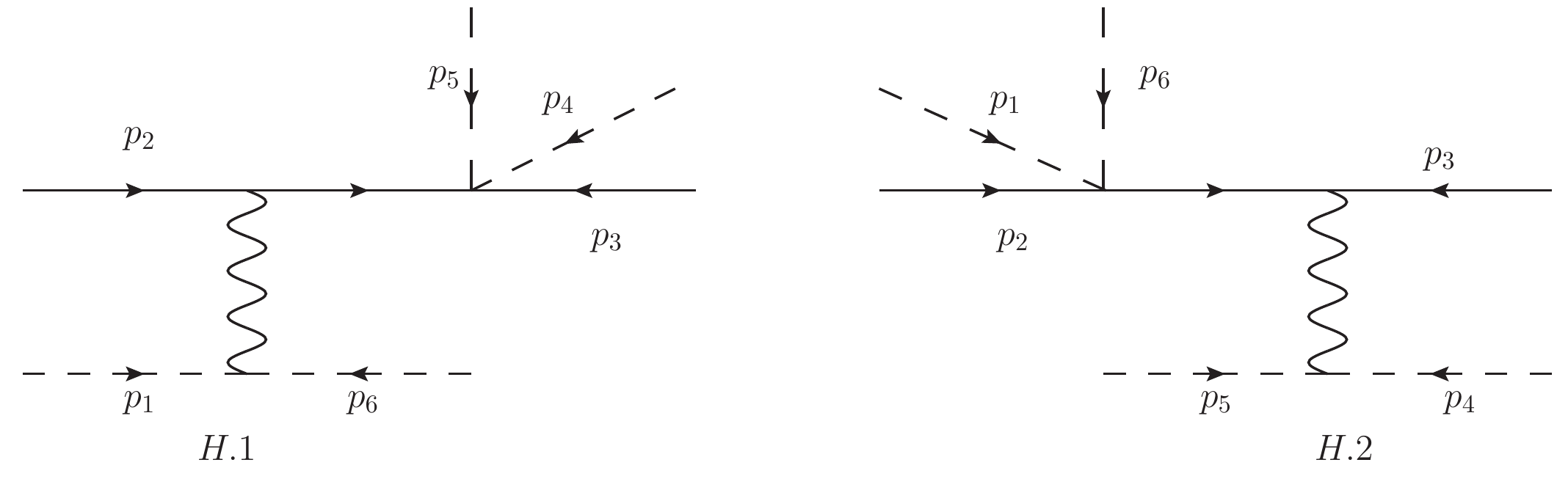}
\end{minipage}
\caption{Feynman diagrams for the amplitude $(\bph(p_1)\ps(p_2))(\bps(p_3)\ph(p_4))(\bph(p_5)\ph(p_6))$ \label{figfy}}
 \end{figure}

In this section, we compute the six point scattering amplitude $\qty(\bph(p_1)\ps(p_2))\qty(\bps(p_3)\ph(p_4))\qty(\bph(p_5)\ph(p_6))$ and demonstrate that it is well behaved under the BCFW deformations. The  Feynman diagrams that contribute to the six point function under consideration are displayed in fig \ref{figfy}.

\begin{figure}[h]
\centering
\begin{minipage}{\textwidth}
\includegraphics[width=14cm,height=4.5cm]{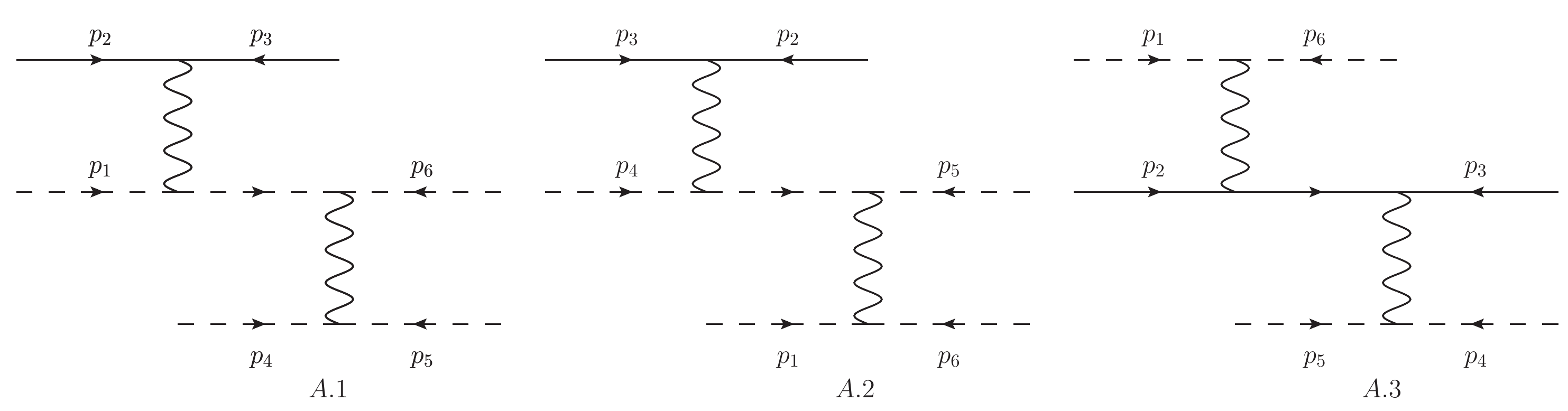}
\end{minipage}
\begin{minipage}{\textwidth}
\includegraphics[width=12cm,height=4cm]{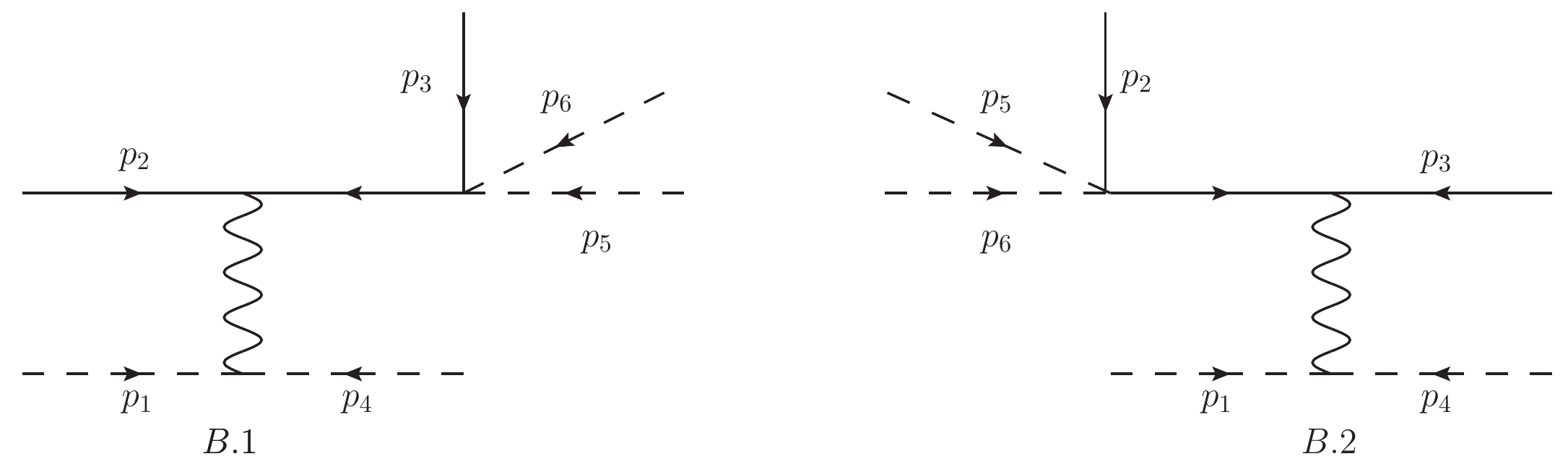}
\end{minipage}
\begin{minipage}{\textwidth}
\includegraphics[width=11cm,height=4cm]{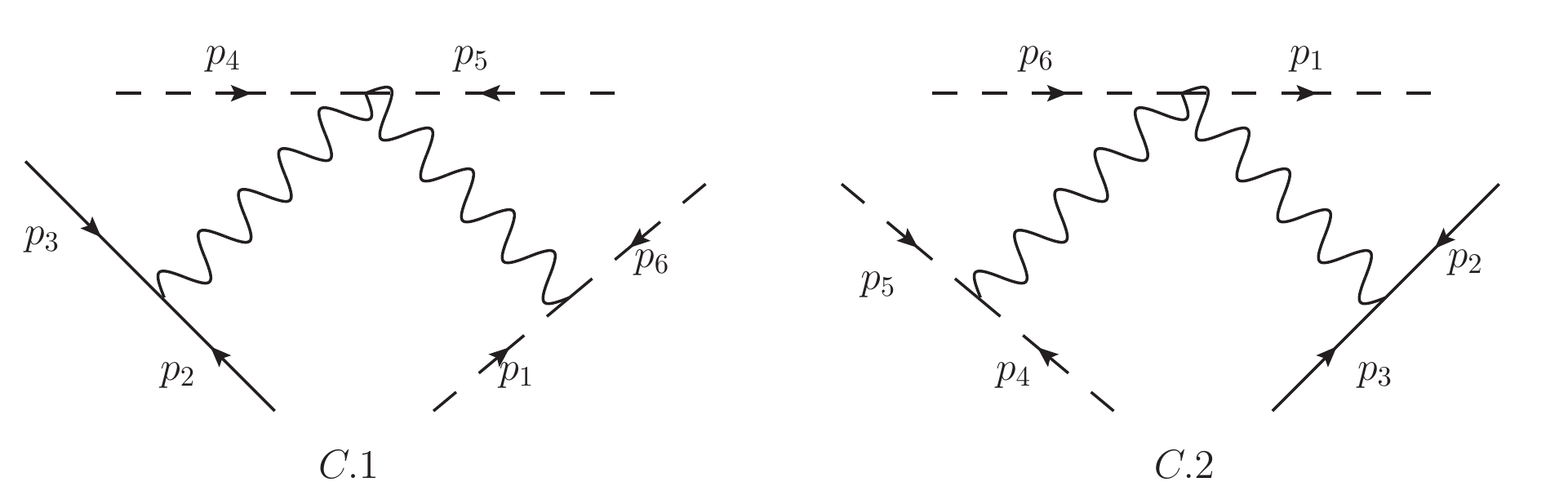}
\end{minipage}
\begin{minipage}{\textwidth}
\includegraphics[width=11cm,height=3.5cm]{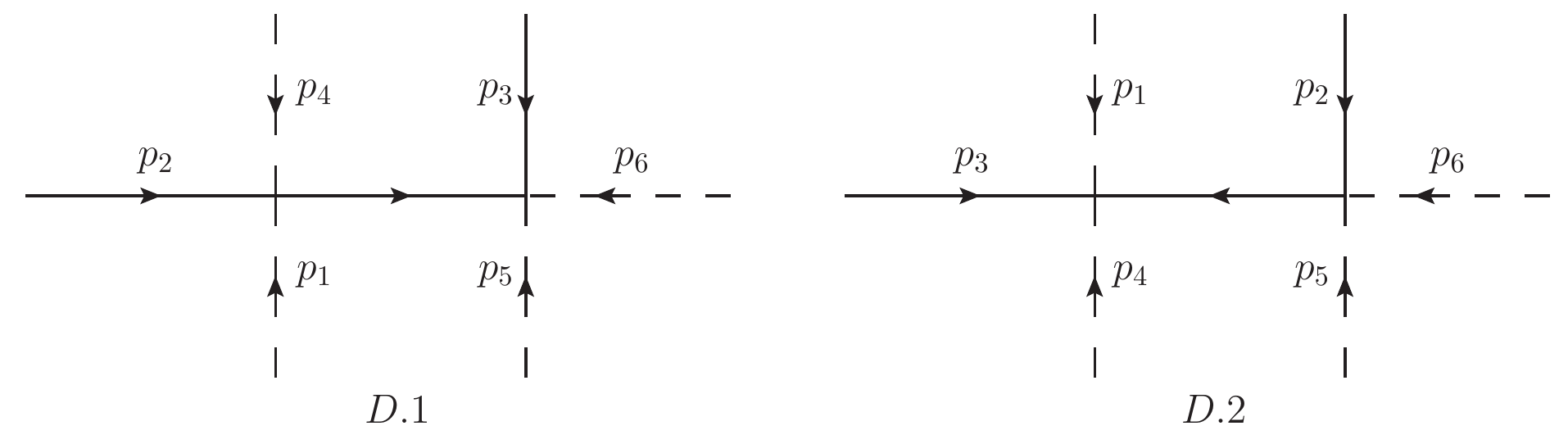}
\end{minipage}
\begin{minipage}{\textwidth}
\includegraphics[width=5.4cm,height=4cm]{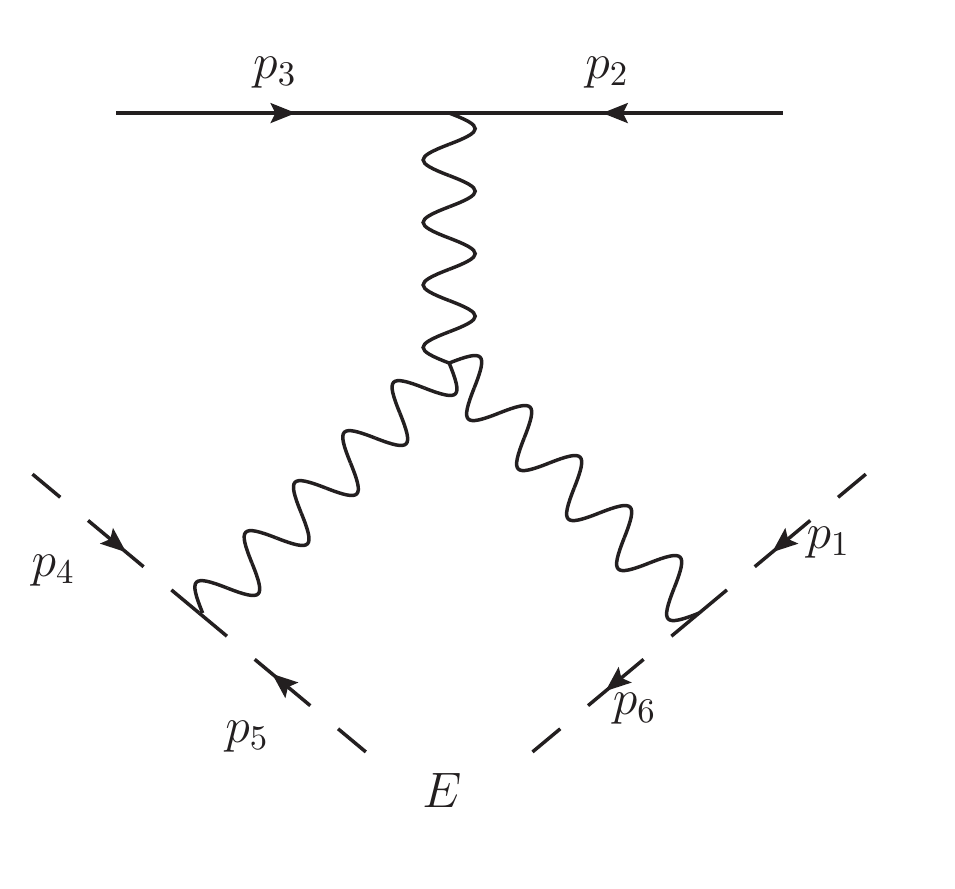}
\end{minipage}
\end{figure}
 We give below explicit expression for each diagram appearing in Fig.\ref{figfy}
\begin{align}
 A_1 = & -\f{16\pi^2i}{\ka^2} \f{p_1.(p_2-p_3)}{p_{23}^2 p_{45}^2 p_{123}^2} \lan 23 \ran \lan 45 \ran \lan 56 \ran \lan 46 \ran \label{A1}\\
  A_2 = & \f{16\pi^2i}{\ka^2} \f{p_4.(p_2-p_3)}{p_{23}^2 p_{16}^2 p_{234}^2} \lan 23 \ran \lan 16 \ran \lan 65 \ran \lan 15 \ran \label{A2}\\
 A_3 = & \f{4i\pi^2}{\ka^2} \f{\lan 16 \ran \lan 45 \ran }{p_{16}^2 p^2_{45} p^2_{126}}\left(\lan 21\ran (\lan 34\ran \lan 5|p_{12}|6\ran +\lan 35\ran \lan 6 |p_{12}|4\ran)+ \lan 26\ran (\lan 34\ran \lan 1|p_{26}|5\ran+\lan 35\ran \lan 1|p_{26}|4\ran)\right) \label{A3} \\
 B_1= &\f{8\pi^2i}{\ka^2} \lan 14\ran \left(\f{\lan 1| p_{56}|3 \ran \lan 24\ran +\lan 3| p_{56}|4\ran \lan 21\ran}{p_{14}^2 p_{356}^2}\right) \label{B1}\\
 B_2 = & \f{8\pi^2i}{\ka^2} \lan 14\ran \left(\f{\lan 1| p_{56}|2 \ran \lan 34\ran +\lan 2| p_{56}|4\ran \lan 31\ran}{p_{14}^2 p_{256}^2}\right)\label{B2}\\
 C_1= & -\f{8\pi^2i}{\ka^2}\left(\f{\lan 2|p_1|3\ran (p_{23}.p_6)-\lan2| p_6|3\ran (p_{23}.p_1)}{p^2_{23} p^2_{16}}\right) \label{C1}\\
 C_2= & -\f{8\pi^2i}{\ka^2}\left(\f{\lan 2|p_5|3\ran (p_{23}.p_4)-\lan2| p_4|3\ran (p_{23}.p_5)}{p^2_{23} p^2_{45}}\right)\label{C2}\\
 D_1 = & \f{8\pi^2 i}{\ka^2} \f{\lan 2| p_{14}|3\ran}{p_{124}^2}\label{D1}\\
D_2 = & \f{8\pi^2 i}{\ka^2} \f{\lan 2| p_{56}|3\ran}{p_{256}^2}\label{D2}\\
E = &-\f{4\pi^2i}{\ka^2}\lan45\ran \lan 23\ran \left( \f{(\lan12\ran^2-\lan13\ran^2) \lan 46\ran\lan 56\ran - (\lan 26\ran^2-\lan 36\ran^2)\lan 14\ran \lan 15\ran}{p_{23}^2 p_{45}^2 p_{16}^2} \right) \label{E}\\
F_1 = &\f{8\pi^2i}{\ka^2} \f{\lan 23\ran \lan 16\ran \lan 65\ran \lan 15\ran}{p^2_{16} p^2_{234}} \label{F1}\\
F_2= &\f{8\pi^2i}{\ka^2} \f{\lan 23\ran \lan 45\ran \lan 56\ran \lan 46\ran}{p^2_{45} p^2_{231}} \label{F2}\\
G = & \f{4\pi^2 i}{\ka^2} \f{\lan 2| p_{16}|3\ran}{p_{126}^2}\label{G}\\
H_1= & \f{4\pi^2 i}{\ka^2}\lan 16\ran\left(\f{\lan 12\ran \lan 6|p_{12} |3 \ran -\lan 26\ran \lan 1 |p_{26}| 3\ran}{p_{16}^2 p_{126}^2}\right) \label{H1}\\
H_2 =& -\f{4\pi^2i}{\ka^2}\lan45\ran \left(\f{\lan35\ran \lan 2|p_{16}|4\ran+\lan34\ran \lan2|p_{16}|5\ran}{p^2_{45}p^2_{126}}\right)
\label{H2}
\end{align}
It is easy to verify that the the asymptotic behavior of the full set of diagram is well behaved by deforming the momentum $p_1$ and $p_2$, as discussed in section \ref{higherpt} and \ref{bacg}. We apply the BCFW deformations in the large z limit using
\begin{align}
 p_2 &\to q z^2 \\
 p_1 &\to- q z^2 \  
 \end{align}
and obtain the asymptotic behavior of the amplitudes to leading order in $z$ as follows. The diagrams $A_2,C_1,D_2,F_1$ in \eqref{figfy} go as $O(1/z)$ to the leading order in the large z limit.
\begin{align}
A_2\sim& \f{2\pi^2i}{\ka^2 z} \f{(q.p_4)\lan q3\ran \lan q6\ran \lan 65\ran \lan q5\ran}{(q.p_3)(q.p_{34})(q.p_6)}+\mathcal O \qty(\frac1{z^3})\\
B_2\sim&  \f{2\pi^2i}{\ka^2 z} \f{\lan q4\ran (\lan q|p_{56}|q\ran \lan 34\ran +\lan q |p_{56}|4\ran \lan 3q\ran )}{(q.p_4) (q.p_{56})}+\mathcal O \qty(\frac1{z^3})\\
C_1\sim& \f{2\pi^2 i}{\ka^2 z}\f{\lan q|p_6|3\ran}{(q.p_6)}+\mathcal O \qty(\frac1{z^3})\\
D_2 \sim &\f{4\pi^2 i }{\ka^2 z} \f{\lan q|p_{56}|3\ran }{q.p_{56}}+\mathcal O \qty(\frac1{z^3})\\
F_1\sim &\f{2\pi^2 i}{\ka^2 z} \f{\lan q3\ran \lan q6\ran \lan 65\ran \lan q5\ran }{(q.p_6) (q.p_{34})}+\mathcal O \qty(\frac1{z^3})\\
\end{align}
For the remaining diagrams we just display the leading large $z$ behavior. They are given by 
\begin{align}
&A_1 \sim -\f{8\pi^2i z}{\ka^2} \f{\lan q3\ran \lan 45 \ran \lan 56\ran \lan 46\ran}{p_{45}^2 p^2_{123}}+{\mathcal O}\qty(\frac1z) && F_2 \sim \f{8\pi^2i z}{\ka^2} \f{\lan q3\ran \lan 45 \ran \lan 56\ran \lan 46\ran}{p_{45}^2 p^2_{123}}+{\mathcal O}\qty(\frac1z)\\
&B_1\sim \f{8\pi^2 i  z}{\ka^2} \f{\lan q| p_{56}|3\ran}{p_{356}^2}+{\mathcal O}\qty(\frac1z) \ && \ D_1\sim -\f{8\pi^2 i  z}{\ka^2} \f{\lan q| p_{56}|3\ran}{p_{356}^2}+{\mathcal O}\qty(\frac1z)\\
&A_3\sim -\f{4\pi^2 i z}{\ka^2} \f{\lan 45\ran (\lan 34\ran \lan q|p_{34}|5\ran +\lan 35\ran \lan q|p_{35}|4\ran)}{p^2_{45}p^2_{126}}+{\mathcal O}\qty(\frac1z) && H_2\sim \f{4\pi^2 i z}{\ka^2} \f{\lan 45\ran (\lan 34\ran \lan q|p_{34}|5\ran +\lan 35\ran \lan q|p_{35}|4\ran)}{p^2_{45}p^2_{126}}+{\mathcal O}\qty(\frac1z)\\
&C_2\sim \f{2\pi^2 i z}{\ka^2} \f{\lan q4\ran \lan q5\ran \lan 3q\ran \lan 45\ran }{p^2_{45} (q.p_3)}+{\mathcal O}\qty(\frac1z) &&  E\sim -\f{2\pi^2 i z}{\ka^2} \f{\lan q4\ran \lan q5\ran \lan 3q\ran \lan 45\ran }{p^2_{45} (q.p_3)}+{\mathcal O}\qty(\frac1z)\\
&G\sim-\f{4\pi^2 i z}{\ka^2} \f{\lan q|p_{45}|3\ran }{p^2_{126}}+{\mathcal O}\qty(\frac1z)  &&  H_1 \sim \f{4\pi^2 i z}{\ka^2} \f{\lan q|p_{45}|3\ran }{p^2_{126}} +{\mathcal O}\qty(\frac1z)
\end{align}
Even though some of the individual diagrams are divergent linearly in $z$, the divergences in the total amplitude cancel pair wise in the large $z$ limit as is evident from the way we have written the results. For example linear in $z$ behavior cancelling pair wise in $\left(A_1,F_2\right),~(B_1,D_1)$ etc.
Thus the total amplitude is well behaved as $z\to\infty$. A straightforward computation yields the analogous result for the $z\to 0$ limit. Thus the amplitude $A_6\qty((\bph(p_1)\ps(p_2))(\bps(p_3)\ph(p_4))(\bph(p_5)\ph(p_6)))$ is well behaved under the BCFW deformations both at $z\to \infty$ and $z\to 0$. 

Towards the end of \S\ref{recrel} we had mentioned that four of the diagrams are reproduced in the factorization channel $p_f=p_{256}$, these diagrams are $B_1,B_2,D_1,D_2$ in fig \ref{figfy}. The remaining eleven diagrams in fig \ref{figfy} are reproduced in the factorization channel $p_f=p_{234}$.

\subsection{A Dyson-Schwinger equation for all loop  six point correlator}
\begin{figure}[h]
\begin{center}
\includegraphics[width=8cm,height=2.9cm]{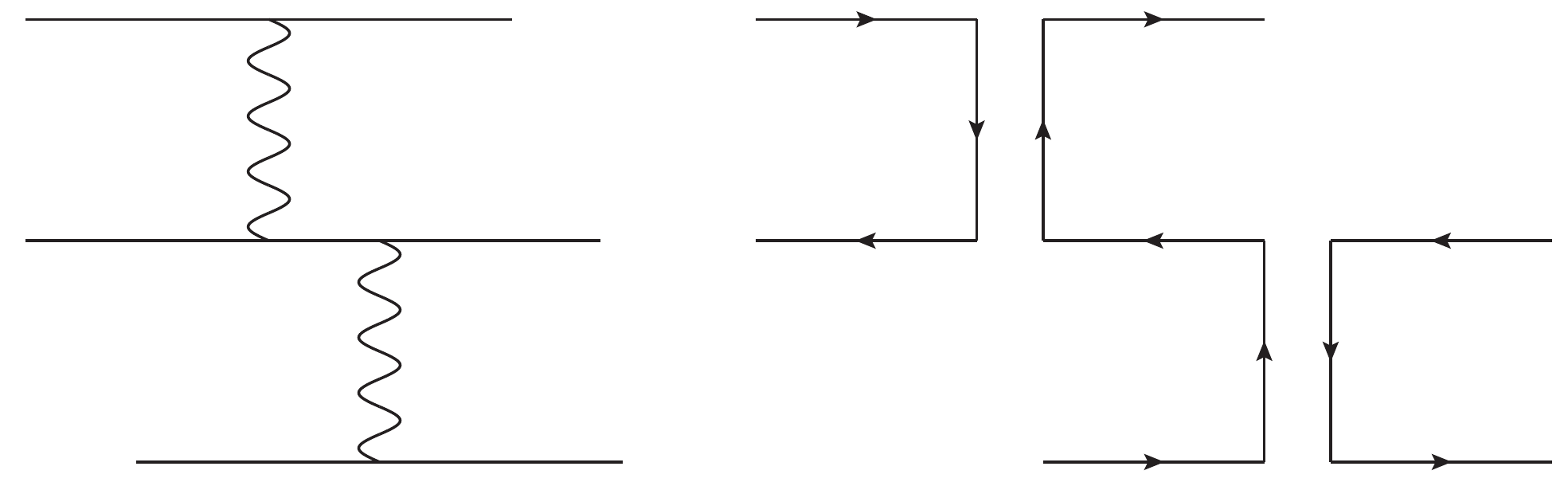}
  \caption{\label{colorst} Six point correlator: We display tree diagrams in supersymmetric light cone gauge. For simplicity we have only displayed the ladder diagrams.The tree diagrams are of order ${\mathcal O}(\f{1}{\ka^2})$ since the gauge field propagator contributes a factor of ${\mathcal O}(\f{1}{\ka})$.}
 \end{center}
\end{figure}

\begin{figure}[b]
\centering
  \begin{minipage}{\textwidth}
\includegraphics[width=10cm,height=8cm]{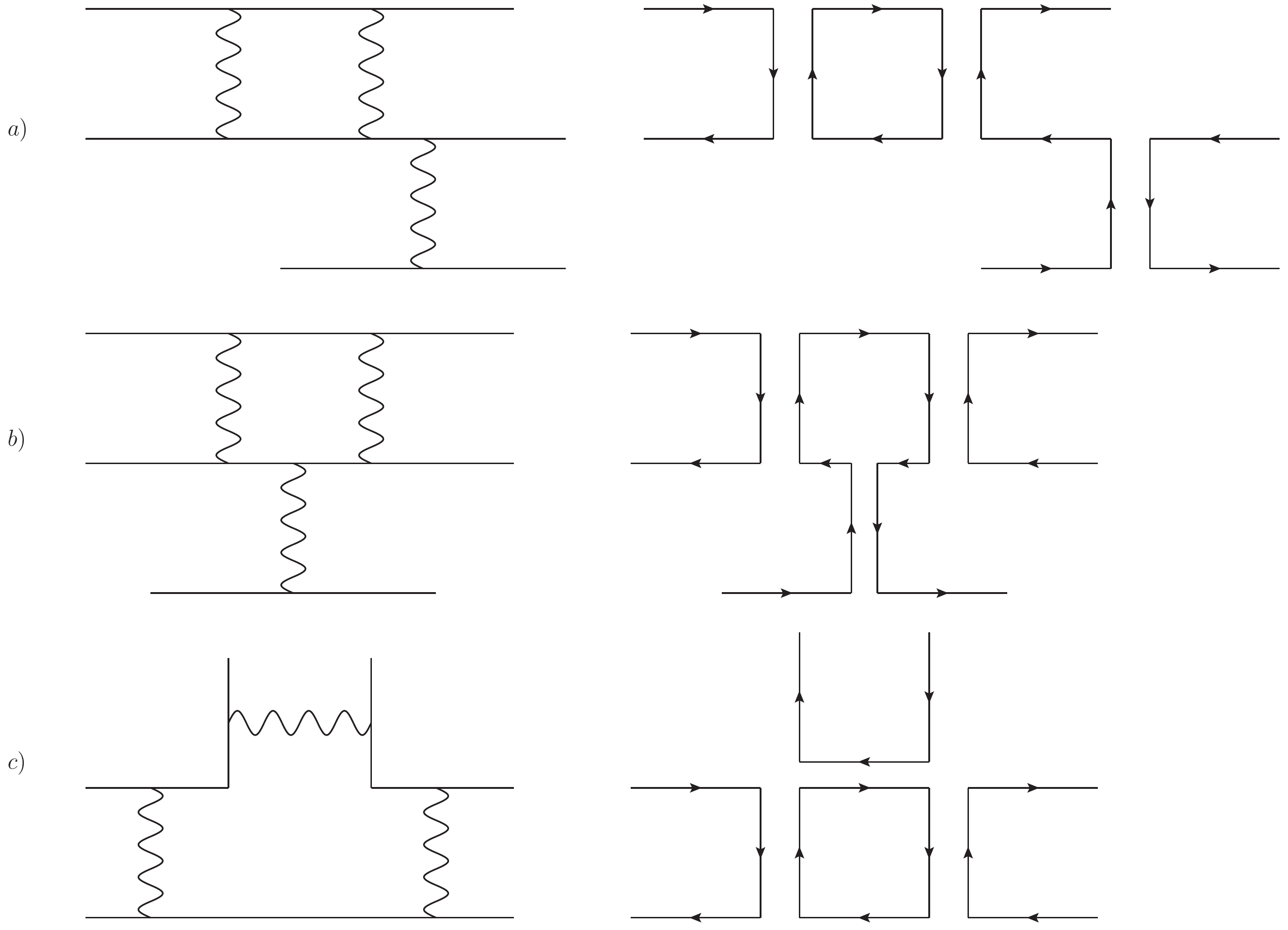}
  \end{minipage}
    \caption{\label{colorst1} Six point correlator: We have listed the various contributions to the one loop correlator in supersymmetric light cone gauge. For simplicity we have displayed only the ladder diagrams. In fig a) and c) the three gauge field propagators contribute a factor of ${\mathcal O}(\f{1}{\ka^3})$ and the single color loop gives a factor of $N$, leading to a contribution of the order 
 $\frac{\la}{\kappa^2}$. Note that this is of the same order in $\ka$ as the tree level diagram displayed in fig \ref{colorst}.  On the other hand fig b) has three gauge fields and no color loops, rendering it to be $O(\f{1}{\ka^3})$.}
\end{figure}

As we saw earlier, the basic building block of higher point amplitudes in the Chern-Simons matter theories at the tree level is the four point amplitude.  In this section we describe the Dyson-Schwinger construction of the all loop six point correlator
\beq\label{sixptcor}
\lan\bPh^i(p+q,\te_1)\Ph_i(-p,\te_2) \bPh^j(k+q',\te_3) \Ph_j(-k-q,\te_6)\bPh^k(r,\te_5)\Ph_k(-r-q',\te_4)\ran
\eeq
using the superspace Schwinger-Dyson construction developed in \cite{Inbasekar:2015tsa}. In the above $\Phi^i$ is a complex scalar superfield in $\mN=1$ superspace defined by
\beq
\Phi^i=\phi^i +\te\ps^i-\te^2 F^i
\eeq
where $\ph^i$ is a complex scalar, $\ps^i$ is a complex fermion and $F^i$ is a complex auxiliary field. The $\mN=2$ theory can be written in $\mN=1$ superspace in terms of $\Phi^i$. For more details see \cite{Inbasekar:2015tsa}. Before presenting the central idea it is informative to understand the color structure of the tree level and one loop amplitudes in the theory. In the supersymmetric Light cone gauge these are described succinctly in fig \ref{colorst} and in fig \ref{colorst1}.
It turns out that, there are six different diagrams for a given color contracted correlator. We have displayed only one in Fig.\ref{colorst} for brevity. 
The situation is a little bit more complicated at one loop as three different type of diagrams can appear as displayed in fig \ref{colorst1}. Note that diagrams like fig \ref{colorst1} b) are  suppressed in the large $N, \ka$ limit (keeping $\lambda=\frac{N}{\kappa}$ fixed). So they don't contribute to the Schwinger-Dyson equation at this order. It can be checked that these type of diagrams continue to remain suppressed at higher loops.

\begin{figure}[h]
\begin{center}
\includegraphics[width=16cm,height=6cm]{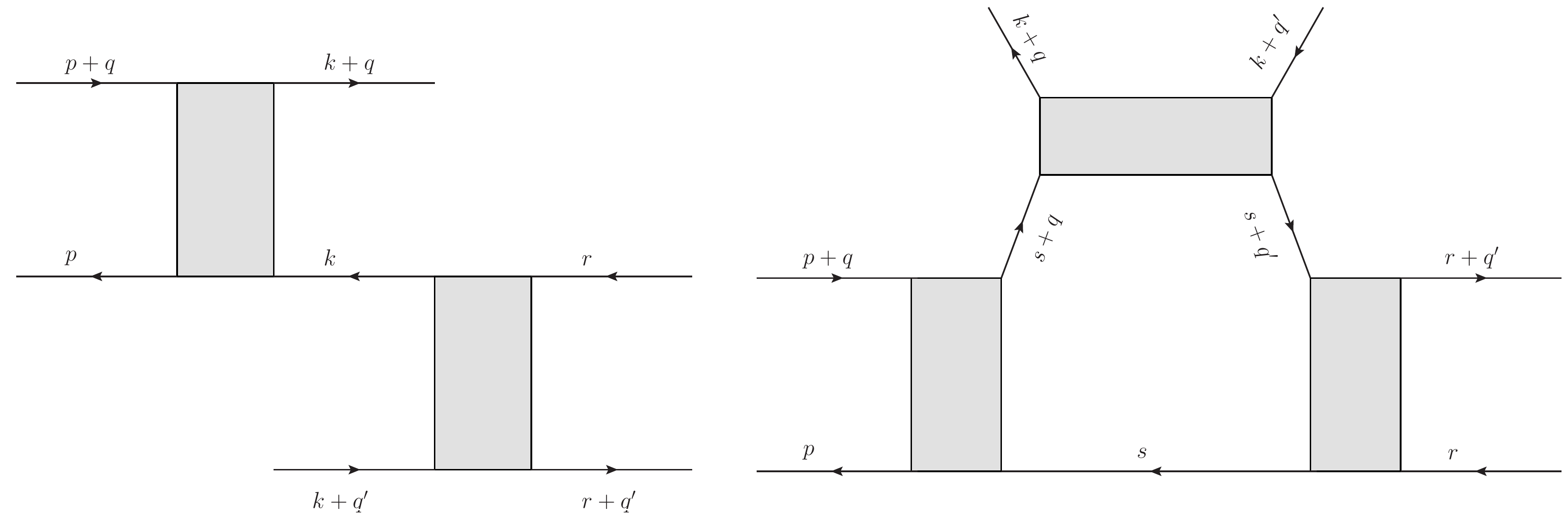}
  \caption{\label{6pttreex} Six point correlator in superspace: The grey boxes represent the $\mN=2$ all-loop four point correlator computed in \cite{Inbasekar:2015tsa}}
 \end{center}
\end{figure}

This paves way for the construction of all loop higher point correlators entirely in terms of all-loop four point correlators at least in the planar approximation. The case for the six point correlator is displayed in (see fig \ref{6pttreex}).
It is straightforward to write down the correlator for the first diagram in \ref{6pttreex}, the second contribution however requires a loop integration over both intermediate grassmann and momentum variables and is quite complicated, we defer a detailed treatment to future works. 

\end{document}